\documentclass[twocolumn,superscriptaddress,floatfix,preprintnumbers,prl]{revtex4}
\usepackage{ORI_Group_style}

\begin{document}

\title{Ultrasensitive Inertial and Force Sensors with Diamagnetically Levitated Magnets}

\author{J. Prat-Camps}
\affiliation{Institute for Quantum Optics and Quantum Information of the
Austrian Academy of Sciences, A-6020 Innsbruck, Austria.}
\affiliation{Institute for Theoretical Physics, University of Innsbruck, A-6020 Innsbruck, Austria.}

\author{C. Teo}
\affiliation{Institute for Quantum Optics and Quantum Information of the
Austrian Academy of Sciences, A-6020 Innsbruck, Austria.}
\affiliation{Institute for Theoretical Physics, University of Innsbruck, A-6020 Innsbruck, Austria.}
\affiliation{Centre for Bio-imaging Sciences, National University of Singapore, 14 Science Drive 4, 117557, Singapore.}

\author{C.~C. Rusconi}
\affiliation{Institute for Quantum Optics and Quantum Information of the
Austrian Academy of Sciences, A-6020 Innsbruck, Austria.}
\affiliation{Institute for Theoretical Physics, University of Innsbruck, A-6020 Innsbruck, Austria.}

\author{W. Wieczorek}
\affiliation{Vienna Center for Quantum Science and Technology (VCQ), Faculty of Physics, University of Vienna, A-1090 Vienna, Austria.}

\author{O. Romero-Isart}  
\affiliation{Institute for Quantum Optics and Quantum Information of the
Austrian Academy of Sciences, A-6020 Innsbruck, Austria.}
\affiliation{Institute for Theoretical Physics, University of Innsbruck, A-6020 Innsbruck, Austria.}

\begin{abstract}
We theoretically show that a magnet can be stably levitated on top of a punctured superconductor sheet in the Meissner state without applying any external field. The trapping potential  created by such induced-only superconducting currents is characterized for magnetic spheres ranging from tens of nanometers to tens of millimeters. Such a diamagnetically levitated magnet  is predicted to be extremely well isolated from the environment. We therefore propose to use it as an ultrasensitive force and inertial sensor. A magnetomechanical read-out of its displacement can be performed by using superconducting quantum interference devices. An analysis using current technology shows that force and acceleration sensitivities on the order of $10^{-23}\text{N}/\sqrt{\text{Hz}}$ (for a 100 nm magnet) and $10^{-14}g/\sqrt{\text{Hz}}$ (for a 10 mm magnet) might be within reach in a cryogenic environment.
Such unprecedented sensitivities can be used for a variety of purposes, from designing ultra-sensitive inertial sensors for technological applications (\eg~gravimetry, avionics, and space industry), to scientific investigations on measuring Casimir forces of magnetic origin and gravitational physics. 
\end{abstract}

\maketitle

Most modern force and inertial sensors are based on the response of a mechanical oscillator to an external perturbation. Such sensors find applications in a wide range of domains: from measuring accelerations in smartphones and automobiles~\cite{Bogue2013} in present-day technology, to being used on the cutting edge of research for magnetic resonance force microscopy~\cite{Rugar2004,Degen2009, bachtold2013}, mass spectroscopy at the single-molecule level~\cite{Naik2009}, and measuring gravitational and Casimir physics at short distances~\cite{Geraci2008,Geraci2010,ArkaniHamed1998,BookCasimir,Klimchitskaya2009}. Most force and inertial sensors are based on microfabricated clamped mechanical oscillators, whose sensitivity is ultimately limited by mechanical dissipation due to material and clamping losses \cite{imboden_dissipation_2014}. Levitation offers a clear route to avoiding these loss mechanisms. Indeed, the most precise commercial accelerometers are based on levitated systems: the superconducting gravimeter, which levitates a superconducting centimeter-sized sphere in the mixed superconducting state to achieve acceleration sensitivities of $3.1\times10^{-10} g/\sqrt{\rm Hz}$~\cite{goodkind1999}, and the MicroStar accelerometer, which electrostatically levitates a centimeter-sized cube in space leading to $10^{-11} g/\sqrt{\text{Hz}}$~\cite{christophe2015}. In research, different levitated systems are being explored to push into unexplored levels of sensitivity. This includes the demonstration of a record force sensitivity of $4\times10^{-22}$N/$\sqrt{\rm Hz}$ with an ion crystal~\cite{Biercuk2010}, the use of optically levitated dielectric nanospheres~\cite{ORI2010, Chang2010, Barker2010,Raizen2010,Gieseler2012,Kiesel2013,Millen2015} as novel force sensors with promising sensitivities~\cite{Yin2013,Rodenburg2016,Ranjit2016} of $2\times 10^{-20}~\text{N}/\sqrt{\text{Hz}}$~\cite{Gieseler2013}, and matter-wave interferometry using clouds of atoms with a sensitivity of $\sim10^{-9} g/\sqrt{\rm Hz}$~\cite{Hu2013, Abend2016}.

In this Letter, we aim at exploiting the exquisite isolation from the environment provided by magnetic levitation in a cryogenic environment. In particular, we propose an all-magnetic passively-levitated sensor that can be scaled over a broad range of sizes and is predicted to  reach unprecedented ultra-high force and inertial sensitivities of $10^{-23}\text{N}/\sqrt{\text{Hz}}$ and $10^{-14}g/\sqrt{\text{Hz}}$, respectively. We show that a spherical particle with a permanent magnetic moment can be stably trapped on top of a punctured superconducting (SC) plane in the Meissner state, without the application of external magnetic fields.
The hole in the SC surface introduces an effective pinning center that, together with the gravitational force, confines the magnet in three dimensions.
Since diamagnetic levitation due to superconductivity does not have any associated length scale, as opposed to the light's wavelength in optical levitation~\cite{Pflanzer2012,Jain2016}, it can be applied to magnets of any size as long as fields in the SC do not prevent superconductivity. The SC surface in the Meissner state ({\em i.e.}~without superconducting vortices) provides a general lossless levitation mechanism. Furthermore, low frequency magnetic field fluctuations arising from the surface are predicted to be minimized in the Meissner state~\cite{Skagerstam06,Hohenester07}. 
The position of the magnet can be precisely measured by placing an array of superconducting interference devices (SQUIDs) in the vicinity of the trap center. The displacement of the magnet couples inductively to the SQUIDs. We shall argue below that these features lead to an alternative approach for ultra-sensitive force and inertial sensing.

\begin{figure}[t]
\begin{center}
\includegraphics[width=0.50\textwidth]{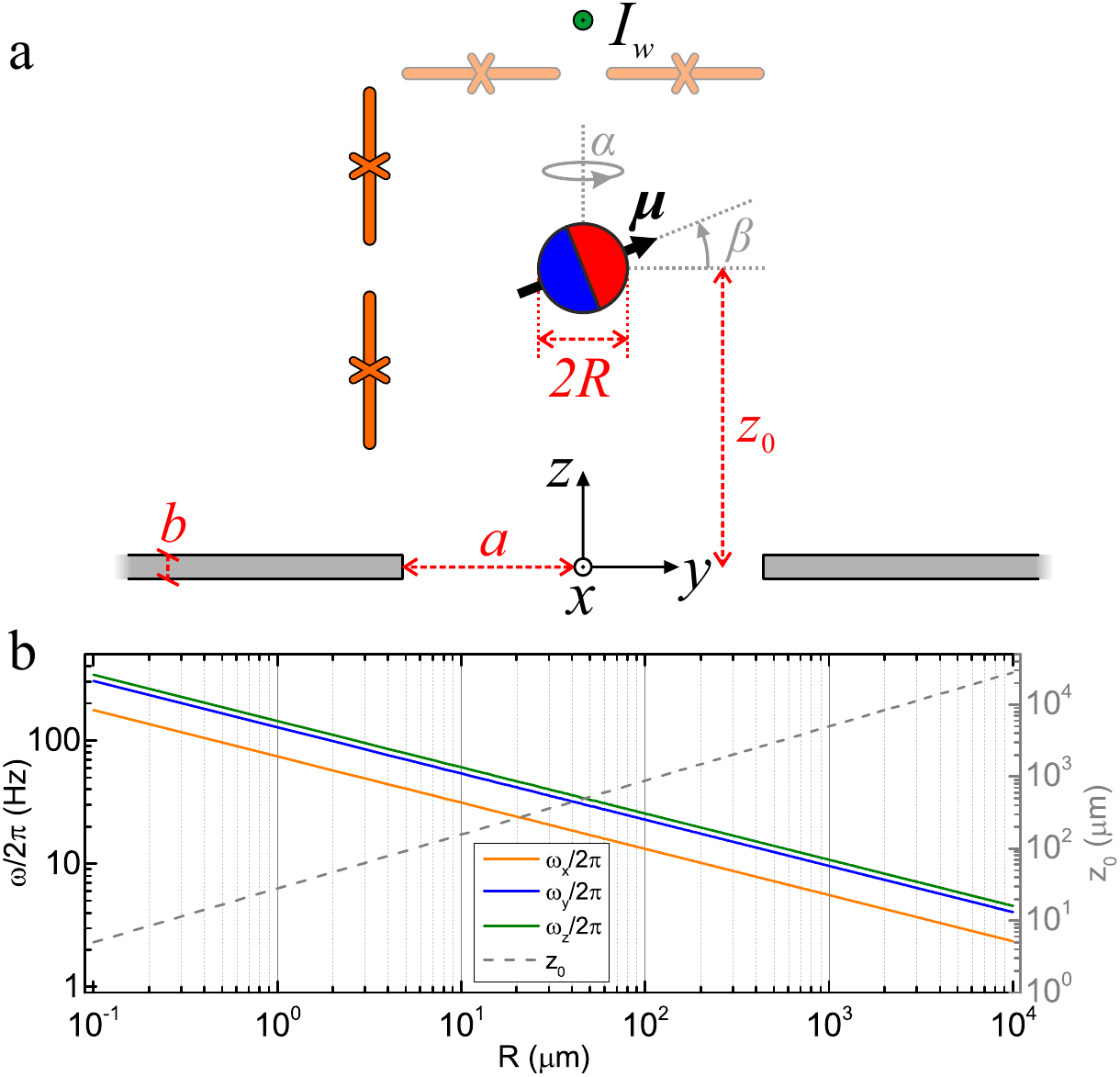}
\caption{ (a) Sketch of the proposal (not to scale). 
(b) Calculated trapping frequencies $\omega_i$ (solid lines, left axis) and trapping distance $z_0$ (dashed line, right axis) as a function of the radius of the magnet, setting $z_0/a=1.8$ (where $a$ depends on $R$) and using the material parameters of ${\rm Nd}_2{\rm Fe}_{14}$B.}
\label{fig.1}
\end{center}
\end{figure}

Let us consider an infinite SC thin film with a circular hole of radius $a$ (whose center defines the origin of coordinates) and thickness $b \ll a$. A  spherical magnet with
radius $R$ and magnetic moment $\boldsymbol{\mu} \equiv \mu(\cos \alpha \cos \beta\,\mathbf{e}_x+\sin \alpha \cos \beta\,\mathbf{e}_y+ \sin \beta \,\mathbf{e}_z)$ is situated on top, see Fig.~\ref{fig.1}(a). 
The SC is described by the London model, which is valid under the approximation that the coherence length of the SC, $\xi$, is much smaller than its London penetration depth, $\lambda$ ($\lambda \gg \xi$). %
We assume a thin film, $b\ll \lambda$, and define the 
two-dimensional Pearl screening length, $\Lambda \equiv 2 \lambda^2/b$ \cite{clem,pearl}.  
The SC is assumed to be in the complete shielding state, namely $\Lambda/a\ll1$~\cite{Brojeny2003,Mawatari2012}.
Importantly, we consider that the SC has been cooled in the absence of any external field, namely that no flux is trapped in the hole. In this case, the SC sheet-current density, ${\bf K}^\text{SC}$, can be calculated from the London equation as ${\bf K}^\text{SC} = - 2 \AB/(\mu_0 \Lambda)$,
where $\mathbf{A}$ is the total magnetic vector potential in the London gauge. 
Zero-field cooling imposes that the fluxoid \cite{tinkham,cardwell} $\Phi'=\Phi + \mu_0 \Lambda \oint_{\rm C}  \mathbf{K}^{\rm SC} \text{d}l /2 $ is zero for any closed path in the SC, including those enclosing the hole ($\Phi$ is the external magnetic flux crossing the surface defined by the closed path C). 
${\bf K}^\text{SC}$ is obtained by making a quasi-static approximation assuming that the SC responds on a timescale much faster than the motion of the magnet. This allows us to numerically solve the 3D magnetostatic problem using a finite-element method with the COMSOL Multiphysics software.  

The magnetic potential felt by the magnet is approximated by
$V_{\rm m}=-\boldsymbol{\mu} \cdot \mathbf{B}_{\rm ind}(\rr)/2$, where $\mathbf{B}_{\rm ind}$ is the field generated by  $\mathbf{K}^{\rm SC}$. This assumes $\mathbf{B}_\text{ind}$ to be sufficiently homogeneous within the volume of the sphere~\footnote{A thorough analysis of the validity of this approximation together with the investigation of the mass limits in diamagnetic levitation in the Meissner state will be addressed elsewhere.}. We remark that the micromagnetic origin of magnetization depends on the size of the magnet. Magnets smaller than a characteristic size, namely the single-domain radius $R_\text{sd}$, consist of a single magnetic domain. Below the so-called blocking temperature, which is the case in a cryogenic environment, the domain is fixed to a given direction.  
Magnets bigger than $R_\text{sd}$ have numerous domains and while their micromagnetic description is cumbersome, can be macroscopically characterized through the hysteresis loop. In that case, the magnet is assumed to be in remanence. We consider magnets made of ${\rm Nd}_2{\rm Fe}_{14}$B, for which $R_\text{sd}\approx 110$nm~\cite{coey}.

The total potential in the presence of gravity reads $V =  V_m + M g z$,
where $M$ is the mass of the  magnet. 
The normalized magnetic potential $\tilde{V}_{\rm m}=V_{\rm m}/V_0$, with $V_0 \equiv \mu_0 \mu^2/(4\pi a^3)$, is numerically calculated as a function of the normalized coordinates, $\tilde{\mathbf{r}} = \mathbf{r}/a$. When the magnetic moment of the magnet is parallel to the SC surface ($\alpha=\pi/2$ and $\beta=0$ such that $\boldsymbol{\mu}=\mu \,\mathbf{e}_y$), it gives rise to a stable trap on the $z$-axis at some $z=z_0$, see further details in Supplemental Material (SM)~\cite{SM}. 
The closest possible trapping point above the SC is at $\tilde{z}_0 \approx 1.168$.
Due to the direction of $\boldsymbol{\mu}$, the trap frequencies in the $x$ and $y$ directions are different and there is a non-negligible cross coupling between $\tilde{y}$ and $\beta$.
The circular hole makes the potential independent on $\alpha$. Alternatively, one could use an ellipsoidal or a polygonal-shaped hole to introduce one or several values of $\alpha$ where energy is minimized. Furthermore, one could consider the use of non-spherical magnets, as recently proposed in the context of magnetometry~\cite{JacksonKimball2016}. For the spherical case, the total potential around the trapping position $\mathbf{r}_0=z_0 \mathbf{e}_z$ and orientation $\alpha=\pi/2$, $\beta=0$, is given by
\begin{align} \label{eq:Potential}
\begin{split}
V(\mathbf{r})\approx & \frac{M}{2}\left(\omega_x^2 x^2 + \omega_y^2 y^2 +\omega_z^2 z^2 \right) +\frac{I}{2}\omega_{\beta}^2 \beta^2+\kappa \beta y.
\end{split}
\end{align}
Here $\mathbf{r}$ is the position vector with origin at $\mathbf{r}_0$,  $\omega_i^2 \equiv \partial_i^2V(0)/M$ (with $i\in\{x,y,z\}$), $\omega_{\beta}^2\equiv \partial_{\beta}^2 V(0)/I$, $\kappa \equiv \partial_{\beta}\partial_y V(0)$, and $I$ is the moment of inertia of the magnet. 
In Fig.~\ref{fig.1}(b) we show the trapping position and frequencies as a function of the radius of the magnet assuming constant mass density and magnetization. Whilst $z_0\sim R^{3/4}$, trapping frequencies show a slow dependence $\sim R^{-3/8}$. Trap depths, defined as the energy (in Kelvins) required to escape the centre of the trap,  grow as $\sim R^{15/4}$ and are of $T\approx 14$K for $R=100$nm. 
The magnetic field at the SC surface is much smaller than the first critical field of Nb (taken as a reference) for all magnet sizes plotted in Fig.~\ref{fig.1}(b). The Euler-Lagrange equations describing the motion of the  magnet~\cite{Rusconi2017} in the potential given by~\eqcite{eq:Potential} can be written in the frequency domain as $\mathbf{X}(\omega)=\boldsymbol{\chi}(\omega)\mathbf{F}(\omega)$, where $\mathbf{X}=(x,y,z,\alpha,\beta)^{\rm T}$ is the vector of coordinates,  $\mathbf{F}=(f_x,f_y,f_z,\tau_{\alpha},\tau_{\beta})^{\rm T}$ is the vector containing external forces ($f$) and torques ($\tau$), and $\boldsymbol{\chi}$ is the susceptibility matrix, see~\cite{SM}.

The position of the magnet can be read out by measuring the magnetic field it creates through a nearby SQUID. 
The flux in the SQUID can be related to the position of the magnet via magnetomechanical coupling factors defined as
$\eta_i\equiv \Phi_0^{-1}\partial_i\Phi(\rr_0)$ (with $i\in\{x,y,z\}$), where $\Phi_0$ is the quantum of flux, and $\Phi(\rr)$ is the flux crossing the SQUID created by the magnet at position $\rr$.
$\eta_i$ depend on the distance, size and arrangement of the SQUID.
In order to measure the three coordinates of the center of mass independently, a suitable arrangement of SQUID loops is used. We consider 4 loops arranged in the same plane, \eg~a plane parallel to XY above the magnet or a plane parallel to XZ, see Fig.~\ref{fig.1}(a).
The position of the magnet can be fully determined through an appropriate linear combination of the flux signals in each loop~\cite{SM}. 

From a practical point of view,
one needs to devise a way to load the magnet and a method to reduce the measurement time of the high-Q oscillator, which is given as a multiple of its ring-down time. 
A possible loading mechanism can rely on guiding the magnet through a conductive cylinder, whose opening is close to the trapping position. Eddy currents induced in the cylinder would slow down the motion of the magnet, which is trapped magnetically upon leaving the cylindrical guide. Reduction of measurement time can be conveniently achieved by feedback cooling, which simultaneously decreases the mechanical quality factor and  the temperature of the oscillator and hence,  maintains a constant overall sensitivity~\cite{Geraci2010}.
 In particular, parametric feedback cooling~\cite{Gieseler2012} could be implemented by applying an external field, such as the one created by an infinite wire with current $I_\text{w}$, parallel to the $x$-axis, passing through the $z$-axis at $z_\text{w}>z_0$, see Fig.~\ref{fig.1}(a). This field modifies the vertical trapping position $z_0$, thereby modulating the trapping frequencies, see~\cite{SM} and~\footnote{A thorough analysis on how to perform feedback cooling in an optimal way such that the added noise does not compromise the overall sensitivity will be addressed elsewhere.}.

The power spectral density (PSD) of a force $F_i$ ($i=\{x,y,z\}$) acting on the magnet, defined as 
$S_{F_i}(\w) = (2 \pi)^{-1} \int_{-\infty}^\infty \avg{F_i(t) F_i(t+\tau)} e^{\im \w \tau} \text{d} \tau$,
is lower bounded by
\be
S_{F_i}(\w)>  S_{F_i}^{\star}(\w) \equiv S_{F_i}^S(\w) + S^N_{F_i}(\w) . \label{eq.sens}
\ee
$S_{F_i}^{\star}(\w)$ is the PSD of the minimal force that can be measured ({\em i.e.}~signal-to-noise ratio of 1), which is limited by contributions due to read-out noise ($S^S_{F_i}$), and to noise forces acting on the magnet ($S^N_{F_i}$). The read-out noise is given by
\be
S_{F_i}^S(\w) = \frac{S_{\Phi}(\w)}{\abs{\chi_{ii}(\w)}^2 \pare{ \Phi_0 \eta_i}^2 },
\ee
where $S_{\Phi}(\w)$ is the PSD describing the flux noise in the SQUID and $\chi_{ii}$ is the diagonal element of the susceptibility matrix $\boldsymbol{\chi}$. 
The contribution $S^N_{F_i}$ contains stochastic forces due to gas collisions and magnetic losses.
The PSDs of accelerations acting on the magnet can be simply obtained as $S_{a_i}=S_{F_i}/M^2$.

We consider the following intrinsic noise sources. 
Regarding the SQUID noise, we assume a low-$T_c$ dc-SQUID mainly affected by white noise~\cite{SQUID_Bible} with a conservative  noise floor of $\sqrt{S_{\Phi}}=10^{-6}\Phi_0 {\rm Hz}^{-1/2}$ for $\mu$SQUIDs \cite{schurig_making_2014}. The noise of an optimized SQUID scales with the self-inductance of the loop, $L$, as $\sqrt{S_{\Phi}}\propto L$~\cite{SQUID_Bible}.
Hence, the noise increases as $\sqrt{S_{\Phi}}\propto s \log (s)$ for bigger SQUIDs, where $s$ is the side length of the SQUID loop. 
 The magnet experiences random gas collision events with a rate proportional to the pressure of the gas, $P$. This gives rise to an effective damping in all coordinates \cite{gamma_gas} approximately given by $\gamma_g \approx 15.8 R^2 P/(M \bar{v}_g)$, where $\bar{v}_g$ is the thermal velocity of the gas molecules. The associated stochastic force PSD, whose expression can be obtained from the fluctuation-dissipation theorem \cite{kuboFDT}, is given by $S_{F}^{\rm g}(\w) = M \gamma_g  k_{\rm B} T/\pi$. 
The magnet fluctuates around its trapping position due to its thermal motion. In the reference frame of the magnet, a time-dependent magnetic field is hence applied. This will cause small fluctuations of the magnetization of the magnet, thereby inducing magnetic losses leading to mechanical damping and a corresponding fluctuating force. In general, one can identify hysteresis losses due to the irreversible relation between the magnetization and the external field as well as eddy-current losses due to  induced currents in the magnet.  
Hysteresis losses can be estimated as follows.
The thermally excited amplitude in each center-of-mass direction is given by $A_i\approx \sqrt{k_{\rm B} T/(M \omega_i^2)}$.  
The field created by the SC currents can be approximated by the one created by an image of $\boldsymbol{\mu}$ at $\mathbf{r}=-z_0\mathbf{e}_z$. The variation of external field at a given point inside the magnet, $\mathbf{r}'$, is 
$\Delta_i \mathbf{H}(\mathbf{r}')=\mathbf{H}(\mathbf{r}'+\mathbf{r}_0+A_i \mathbf{e}_i)- \mathbf{H}(\mathbf{r}'+\mathbf{r}_0)$. The variation of magnetization is 
$\Delta_i \mathbf{M}(\mathbf{r}') \approx \chi_{\rm m} \Delta_i \mathbf{H}(\mathbf{r}')$, where $\chi_{\rm m}$ is the magnetic susceptibility of the magnet in remanence.
The magnetic energy lost per cycle can be estimated as 
$\Delta_i W_{\rm h} \approx \int_V \mu_0 \Delta_i \mathbf{M}(\mathbf{r}')\Delta_i \mathbf{H}(\mathbf{r}')dV'=\mu_0 \chi_{\rm m} \int_V\Delta_i \mathbf{H}(\mathbf{r}')^2 dV'$. This is an overestimation of the hysteresis loss per cycle, both because the three components of the magnetization are assumed to change due to the external field (by using a simple scalar $\chi_{\rm m}$) and because all the energy in the product $\mu_0 \Delta \mathbf{M}\Delta \mathbf{H}$ is considered to be irreversibly dissipated 
\footnote{Taking into account the tiny amplitude of the field oscillations and the coercitive field of the magnet, one can expect a rather linear magnetic behaviour with a narrow hysteresis loop that would contain a small fraction of the estimated energy.}.
The damping rate is then given by $\gamma_{h_i}\approx\omega_i \, \Delta_i W_{\rm h}/(2\pi k_{\rm B} T)$ and, assuming thermal equilibrium, the associated stochastic force is $S_{F_i}^{\rm h}(\w) = M \gamma_{h_i}  k_{\rm B} T/\pi$.
Eddy-current losses can be estimated through a similar procedure. The energy loss per cycle, $\Delta W_{\rm e}$, is proportional to the electrical conductivity of the magnet and the frequency of the field. Considering the poor conductivity of typical magnets, and in particular of ${\rm Nd}_2{\rm Fe}_{14}$B, and the small frequencies involved ($<100$ Hz), one can readily show that $\Delta W_{\rm e}\ll \Delta W_{\rm h}$. 
In the limit of small magnets with a single magnetic domain, the only magnetic dissipative process is related to the alignment of the magnet to a non-parallel external magnetic field, which involves a minimum time scale related to the relaxation of the crystal lattice to the equilibrium orientation (described by the Landau-Lifshitz-Gilbert equation)~\cite{landau35,gilbert2004}. This effect is predicted to be negligible at the low frequencies considered.

\begin{figure}[t]
\begin{center}
\includegraphics[width=0.50\textwidth]{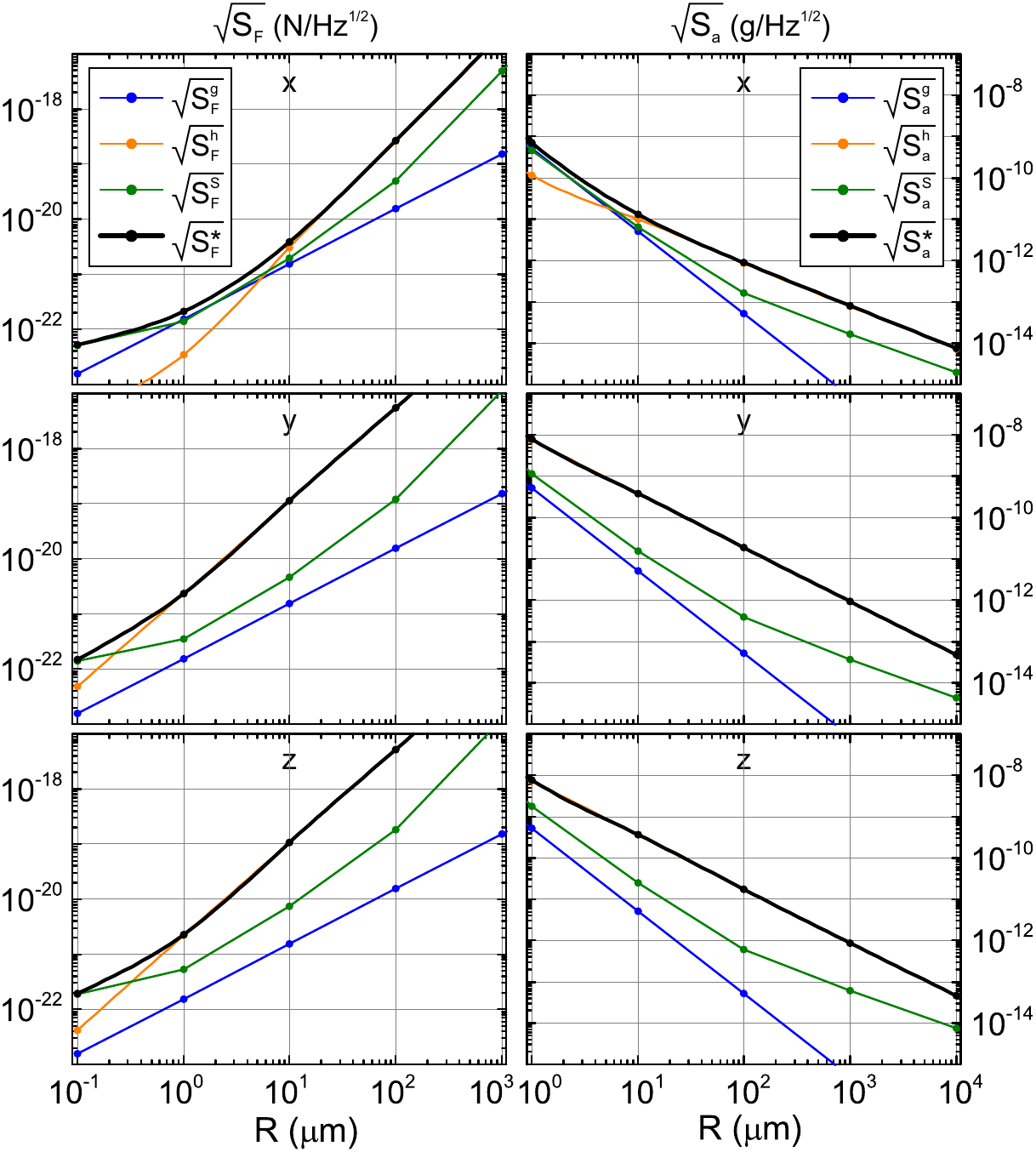}
\caption{Noise calculations as a function of the radius of the magnet (symbols, lines are guides for the eye) in terms of force (left half) and acceleration (right half). Noises are computed at a fraction of the corresponding resonance frequency, see~\cite{SM}.}
\label{fig.2}
\end{center}
\end{figure}

The previous analysis  can now be applied to magnets with sizes spanning over very different scales, from nanometers to millimeters. 
Small masses provide high force sensitivities since the mechanical susceptibility scales as $\chi \propto 1/M$. Force noise due to gas collisions ($\propto R$) and magnetic losses are minimized for small masses. 
On the other limit, large masses provide high sensitivity on the acceleration of the magnet.
Larger magnets create stronger magnetic fields leading to bigger couplings to the SQUIDs. However, losses related to magnetic hysteresis become relevant as the volume of magnetic material increases.
Fig.~\ref{fig.2} shows the noise contributions and the final sensitivity for different sizes of the magnet at the reference temperature of 1K for Nd$_2$Fe$_{14}$B~\cite{coey}, see~\cite{SM}. The largest force (acceleration) sensitivity at small (large) radii is limited by the SQUID noise (hysteresis losses). Recall that hysteresis losses are overestimated, so one could expect even better acceleration sensitivities. 
Sensitivities, evaluated at a fraction of the corresponding resonance frequency, reach $5 \times 10^{-23}$N/$\sqrt{\rm Hz}$ at  $f\sim 18$Hz for a magnet of $R=100$nm (with resonance frequency $f\sim 180$Hz and $Q\sim10^9$) and $7\times10^{-15}$g/$\sqrt{\rm Hz}$ at $f\sim 1$Hz for a magnet of $R=10$mm (resonance frequency $f\sim 2$Hz and $Q\sim10^5$). Such a force sensitivity is more than an order of magnitude better than the current state-of-the art using trapped ions \cite{Biercuk2010}. The acceleration sensitivity is more than three orders of magnitude better than in commercial devices \cite{goodkind1999,christophe2015}.

Such unprecedented sensitivities could be used, among others, to measure inclinations, vibrations, and magnetic field fluctuations. For magnets of $R=10$mm, inclinations on the order of
$7 \times 10^{-15}$rad/$\sqrt{\rm Hz}$
and vibrations on the order of 
$2 \times 10^{-16}$m/$\sqrt{\rm Hz}$ could be detected. Magnetic gradients of up to $5\times10^{-16}$T/(m$\sqrt{\rm Hz}$) at $f\sim 1$Hz would also be detectable.
The latter could be used to detect magnetic fields created by fluctuating currents in nearby solids, \ie~to detect magnetic Casimir forces~\cite{henkel1999}.
For a magnet with $R=10\mu$m close to a silver surface, this force falls within the detectability threshold for separations of up to $\approx15\mu$m  from the surface, see~\cite{SM}.
Electric Casimir forces could be detected by coating the magnet with a non-magnetic dielectric material and approach a dielectric surface to it.  
Further, with appropriate shielding from Casimir forces, the device could also be used to test corrections to the gravitational force at short distances~\cite{Geraci2008,Geraci2010,ArkaniHamed1998}.
A more ambitious goal would be to use the extreme acceleration sensitivity of our device to detect gravitational forces between small masses and accurately characterize Newtons's constant $G$, see~\cite{Schmole2016} and references therein.
Note that the gravitational interaction between a magnet of $R\approx5~\text{mm}$ and another sphere of the same mass separated by a gap of $3~\text{mm}$ could be in principle detected.
Finally, using our device as an inertial sensor could have relevant applications in avionics and space industry. The detection of small variations of gravitation force could also be applied to geological exploration or mining, among others.

In conclusion, we have presented an alternative approach for force and inertial sensing based on diamagnetic levitation of magnets. Remarkably, the concept is rather general and can be applied to magnets with sizes ranging from nanometers to millimeters, spanning over 6 orders of magnitude. The underlying mechanism behind such an astonishing broad window is the diamagnetic levitation provided by the superconductor in the Meissner state. 
The use of a magnet with a strong magnetic moment gives rise to a simple passive trapping scheme, and provides direct ways to read and feedback cool its motion.
Our analysis, including current technologies and realistic assumptions, indicates very promising sensitivities over a wide range of scales, which we hope will motivate its experimental implementation.

This work is supported by the European Research Council (ERC-2013-StG 335489 QSuperMag) and the Austrian Federal Ministry of Science, Research, and Economy (BMWFW). We acknowledge discussions with J. Hofer, G. Kirchmair, C. Navau, A. Sanchez, and C. Schneider.

\newpage

\newpage

\newpage


\section*{SUPPLEMENTAL MATERIAL} 

\section{Numerical characterization of the trapping potential}

\begin{figure}[t]
\begin{center}
\includegraphics[width=0.50\textwidth]{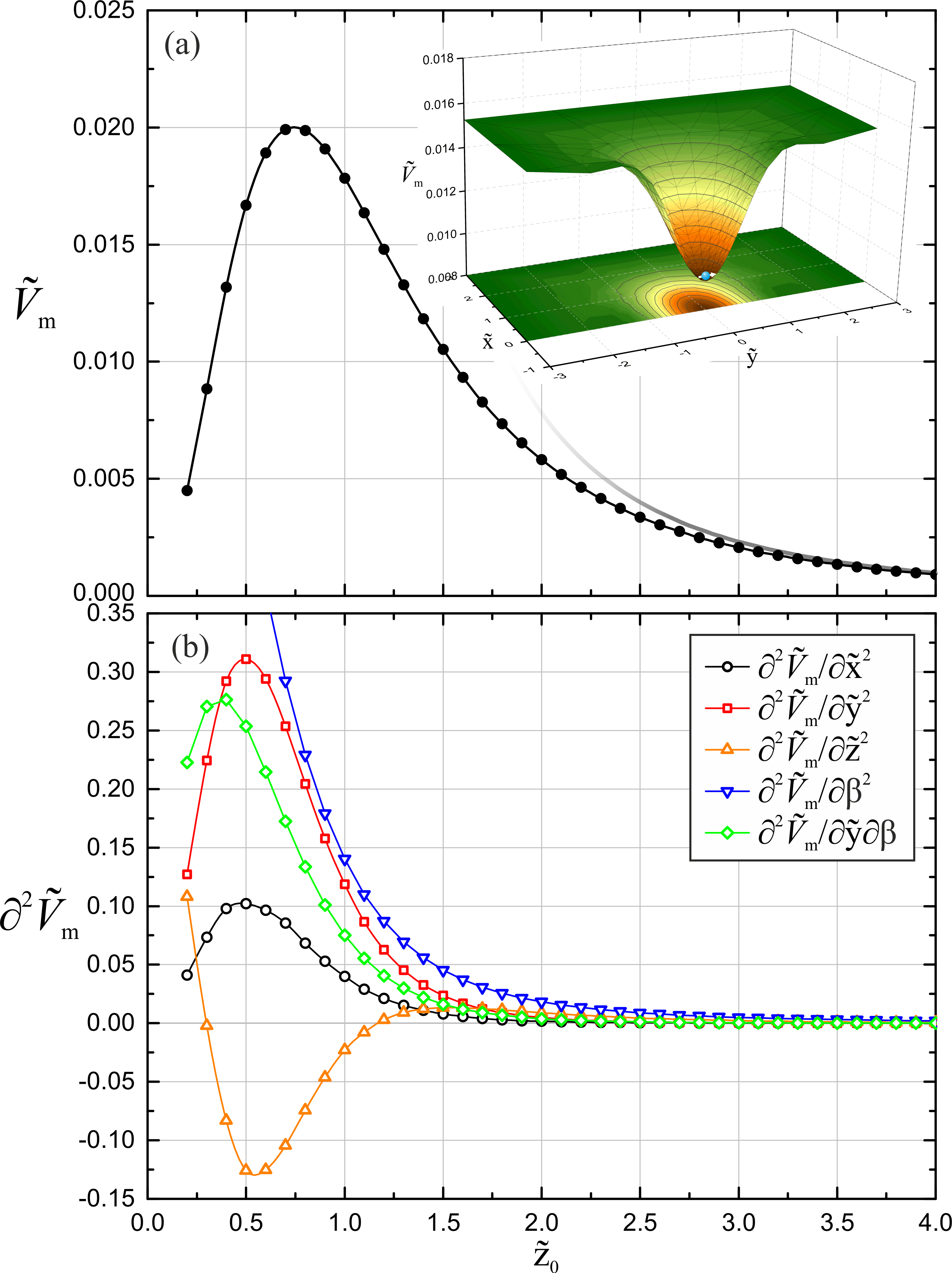}
\caption{ (a) Numerical calculations of $\tilde{V}_{\rm m}$ for a magnet at $\mathbf{r}_0=z_0 \,\mathbf{e}_z$. Gray line is the potential calculated with the image method. Inset shows a surface plot of the magnetic potential for a fixed $\tilde{z}=1.6$.
(b) Second derivatives of the magnetic potential with respect to the spatial coordinates and the angular coordinate $\beta$ evaluated at $\mathbf{r}_0=z_0 \,\mathbf{e}_z$. All calculations consider the SC to be in the ideal complete shielding state ($\Lambda/a\ll1$).}
\label{fig.S1}
\end{center}
\end{figure}

The normalized magnetic potential $\tilde{V}_{\rm m}=V_{\rm m}/V_0$, with $V_0\equiv\mu_0 \mu^2/(4\pi a^3)$, is numerically calculated for a magnet with magnetic moment parallel to the SC surface ($\boldsymbol{\mu}=\mu \,\mathbf{e}_y$). This orientation of the magnet gives rise to a stable trap on the z-axis, as shown in the inset of Fig. \ref{fig.S1}(a), which is plotted in terms of dimensionless coordinates, $\tilde{\mathbf{r}} = \mathbf{r}/a$. 
On the $z$-axis, the potential shows a non-monotonic dependence with $\tilde{z}_0$ [Fig. \ref{fig.S1}(a)]. In the limit $\tilde{z}_0\gg 1$ it agrees, as expected, with the potential created by an infinite ideal superconductor (SC), whose analytical expression obtained from the image method is $\tilde{V}_{\rm m}^{\rm imag}=1/(16\tilde{z}_0^3)$. It can be shown that all second derivatives with respect to crossed spatial coordinates on the $z$-axis are zero, demonstrating that there is no coupling between them. 
Since the magnetic moment is oriented along the $y$-axis, axial symmetry is broken and second derivatives with respect to $\tilde{x}$ and $\tilde{y}$ have different values (see \figref{fig.S1}b). The second derivative with respect to $\tilde{z}$ is zero at $\tilde{z}_0 \approx 1.168$, determining the closest possible trapping point above the SC. Alternatively, one could also trap at $\tilde{z}_0 \approx 0$; in that case gravity would shift the final trapping position slightly below the SC. 
Cross derivatives with respect to $\beta$ show an interesting property of the system; whilst symmetry ensures that $\tilde{x}\beta$ and $\tilde{z}\beta$ derivatives are zero on the $z$-axis, derivatives with respect to $\tilde{y}\beta$ are large. This is also related to the symmetry-breaking direction of the magnetic moment and leads to a coupling between these two coordinates.

\section{Derivation of the equations of motion}

The Lagrangian of the magnet trapped in the potential we characterized reads~\cite{Rusconi2017} 
\begin{align}
\begin{split}
L=&\,\frac{M}{2}\left(\dot{x}^2+\dot{y}^2+\dot{z}^2\right)\\
&+\frac{I}{2}\spare{\dot{\tilde{\alpha}}^2+(\dot{\tilde{\gamma}}-\omega_{\mu})^2+2(\dot{\tilde{\gamma}}-\omega_{\mu})\dot{\tilde{\alpha}}\cos\tilde{\beta}+\dot{\tilde{\beta}}^2 }\\
&-V(\mathbf{r},\beta),
\end{split}
\end{align}
where $V$ is the trapping potential given in Eq.~(1) of the main text, $\tilde{\alpha}$, $\tilde{\beta}$ and $\tilde{\gamma}$ are the Euler angles in the ZYZ convention, and $\omega_{\mu}=\hbar \mu/(I g_e \mu_{\rm B})$, where $g_e$ is the gyromagnetic ratio of the electron. This Lagrangian assumes an ideal hard-magnet with infinite magnetic anisotropy energy such that the magnetic moment is perfectly clamped to the anisotropy axis of the crystal~\cite{Rusconi2017}. In order to express it in terms of the angle $\beta$ defined in the main text, one can make the following change of variables $\tilde{\alpha}=\alpha$, $\tilde{\beta}=\pi/2-\beta$, and $\tilde{\gamma}=\gamma$. The Euler-Lagrange equations are obtained as $d(\partial L/\partial \dot{q})/dt-\partial L/ \partial q=0$ . After linearising them around the trapping position for a non-spinning magnet they read
\begin{align}
\begin{split}
    \ddot{x}+ \omega_x^2 x= & 0,\\
    \ddot{y}+\omega_y^2 y+\kappa \beta/M = & 0,\\
    \ddot{z}+\omega_z^2 z= & 0,\\
    \ddot{\alpha}-\omega_{\mu} \dot{\beta}= & 0,\\
    \ddot{\beta}+\omega_{\beta}^2 \beta +\omega_{\mu} \dot{\alpha}+\kappa y/I= & 0.
\end{split}
\end{align}
All parameters are defined in the main text. One can now introduce fluctuating forces ($f$) and torques ($\tau$) acting on each coordinate, as well as the corresponding damping rates ($\gamma$) assuming the fluctuation-dissipation theorem in thermal equilibrium. Rewriting these equations in the frequency domain one obtains
\begin{align} \label{eq:Help}
\begin{split}
    M\left(\omega_x^2-\omega^2 -i\omega \gamma_x \right) x= & f_x,\\
    M\left(\omega_y^2-\omega^2 -i\omega \gamma_y \right) y+\kappa\, \beta= & f_y,\\
    M\left(\omega_z^2-\omega^2 -i\omega \gamma_z \right) z= & f_z,\\
    I\left(-\omega^2 -i\omega \gamma_{\alpha}\right) \alpha + i\omega\, \omega_{\mu} I\, \beta= & \tau_{\alpha},\\
    \kappa\, y-i\omega\, \omega_{\mu} I\, \alpha +I\left(\omega^2_{\beta}-\omega^2 -i\omega \gamma_{\beta}\right) \beta= & \tau_{\beta}.
\end{split}
\end{align}
This system of linear equations can then be simply solved as 
\begin{equation} \label{eq.xi}
\mathbf{X}(\omega)=\boldsymbol{\chi}(\omega)\mathbf{F}(\omega),
\end{equation}
where $\mathbf{X} \equiv (x,y,z,\alpha,\beta)^{\rm T}$, $\mathbf{F} \equiv (f_x,f_y,f_z,\tau_{\alpha},\tau_{\beta})^{\rm T}$, and $\boldsymbol{\chi}$ is the mechanical susceptibility matrix given by the inverse of the matrix giving the system of linear equations in \eqnref{eq:Help}.

\section{PSD definition, sensitivity and signals analysis}

\subsection{PSD definition}

The power spectral density (PSD) of a variable $A(t)$ is defined as
\be
S_A(\w) \equiv \frac{1}{2 \pi} \int_{-\infty}^\infty  \avg{A(t) A(t+\tau)} e^{\im \w \tau} \text{d} \tau,
\ee
where the autocorrelation function is
\be
 \avg{A(t) A(t+\tau)} = \lim_{T \rightarrow \infty} \inv{T} \int_{-T/2}^{T/2} \text{d}t A(t) A(t+\tau).
\ee

\subsection{Sensitivity}

By considering that the stable trapping point of the magnet is $\mathbf{r}_0$, the flux in the SQUID when the magnet is at $\mathbf{r}_0+\mathbf{r}$ can be approximated to 
\begin{equation}
\Phi(\mathbf{r}_0+\mathbf{r})\approx \Phi(\mathbf{r}_0)+\Phi_0 \eta_x \,x+\Phi_0 \eta_y \,y +\Phi_0 \eta_z \,z, \label{eq.fl}
\end{equation}
where $\mathbf{r}=(x,y,z)$ and $\eta_i\equiv \Phi_0^{-1}\partial_i\Phi(\rr_0)$ for $i=\{x,y,z\}$. Consider the PSD $S_{F_i}^{\rm sig}$ of the force  acting on the magnet that we want to measure. Eq.~\eqref{eq.xi} gives us the position of the magnet as a result of this force. 
The PSD of the position of the magnet as a result of this force signal is $S_{r_i}^{\rm sig}=|\chi_{ii}|^2 S_{F_i}^{\rm sig}$ and, thus, the PSD of the flux signal in the SQUID is
\begin{equation}
S_{\Phi_i}^{\rm sig}=(\Phi_0 \eta_i)^2|\chi_{ii}|^2 S_{F_i}^{\rm sig}.
\end{equation}
There are two types of noise that will limit the sensitivity of this signal; the noise affecting the SQUID (whose PSD is $S_\Phi$) and the noises coming from the stochastic forces acting on the magnet (gas and magnetic, with PSDs $S_{F_i}^{\rm g}$ and $S_{F_i}^{\rm h}$, respectively). The PSD of the position of the magnet due to these stochastic forces is $S_{r_i}^N=|\chi_{ii}|^2(S_{F_i}^{\rm g}+S_{F_i}^{\rm h})$.
The PSD of the flux signal in the SQUID due to all these noises is
\begin{equation}
S_{\Phi_i}^{\rm noise}=S_{\Phi}+(\Phi_0 \eta_i)^2 |\chi_{ii}|^2\left(S_{F_i}^{\rm g}+S_{F_i}^{\rm h}\right).
\end{equation}
The sensitivity condition is given by $S_{\Phi_i}^{\rm sig}>S_{\Phi_i}^{\rm noise}$ which can be rewritten as 
\begin{equation}
S_{F_i}^{\rm sig}>\frac{S_{\Phi}}{|\chi_{ii}|^2(\Phi_0 \eta_i)^2}+(S_{F_i}^{\rm g}+S_{F_i}^{\rm h}).
\end{equation}
This expression corresponds to Eq.~(2) in the main text, where the second term on the right hand side is defined as $S_{F_i}^N$. 

\subsection{Signal analysis}

Apart from the three noise sources analysed in the last section, the system will be affected by other signals such as inclinations and vibrations of the SC sheet as well as magnetic fields produced by nearby objects. Depending on the operating mode of the system, these signals can be considered as part of the noise or, on the contrary, they can be the signals one is interested to measure. 
In this section we analyse the signals produced by inclinations, vibrations of the SC sheet, and magnetic fields.  

\vspace{2mm}

\textbf{\textit{Inclinations}}.
Inclinations of the SC surface with an angle $\Delta \gamma$ around the $y$-axis result in a force $F_x=Mg\,\Delta \gamma$ which Eq.~\eqref{eq.xi} converts into a position signal as $S_{x}^{\rm incl}=|\chi_{xx}M g|^2 S_{\Delta \gamma}$. This position signal can also be interpreted as a result of a force signal such that $S_{F_x}^{\rm incl}=S_{x}^{\rm incl}/|\chi_{xx}|^2$ so  
\begin{equation}
S_{F_x}^{\rm incl}=|M g|^2 S_{\Delta \gamma}.
\end{equation}
Inclinations with an angle $\Delta \beta$ around the $x$-axis result in a force $F_y=(Mg+\kappa)\Delta \beta$ and a torque $\tau_{\beta}=I\omega_{\beta}^2\Delta \beta$. Using Eq.~\eqref{eq.xi}, we find that they only couple to the $y$-coordinate of the magnet, so the PSD of the position signal is $S_{y}^{\rm incl}=|\chi_{yy}(M g+\kappa)+\chi_{y\beta}I\omega_{\beta}^2|^2 S_{\Delta \beta}$. The force signal corresponding to it is $S_{F_y}^{\rm incl}=S_{y}^{\rm incl}/|\chi_{yy}|^2$, namely
\begin{equation}
S_{F_y}^{\rm incl}=\left|\frac{\chi_{yy}(M g+\kappa)+\chi_{y\beta}I\omega_{\beta}^2}{\chi_{yy}} \right|^2 S_{\Delta \beta}.
\end{equation}

\vspace{2mm}

\textbf{\textit{Vibrations}}.
In general, vibrations of the SC surface result in forces and torques acting on the magnet, which are converted into position signals through Eq.~\eqref{eq.xi}. The PSDs of position signals read 
$S_x^{\rm vib}=|\chi_{xx}M\omega_x^2+1|^2 S_{\Delta x}$,
$S_y^{\rm vib}=|\chi_{yy}M\omega_y^2+\chi_{y\beta}\kappa+1|^2 S_{\Delta y}$, and
$S_z^{\rm vib}=|\chi_{zz}M\omega_z^2+1|^2 S_{\Delta z}$. 
The last term in these expressions accounts for the change of distance between the readout system and the magnet as a result of the vibration. The corresponding force signals are
\begin{align}
S_{F_x}^{\rm vib}&=\left|\frac{\chi_{xx}M\omega_x^2+1}{\chi_{xx}}\right|^2 S_{\Delta x},\\
S_{F_y}^{\rm vib}&=\left|\frac{\chi_{yy}M\omega_y^2+\chi_{y\beta}\kappa+1}{\chi_{yy}}\right|^2 S_{\Delta y},\\
S_{F_z}^{\rm vib}&=\left|\frac{\chi_{zz}M\omega_z^2+1}{\chi_{zz}}\right|^2 S_{\Delta z}.
\end{align}

\vspace{2mm}

\begin{figure}[t]
\begin{center}
\includegraphics[width=0.5\textwidth]{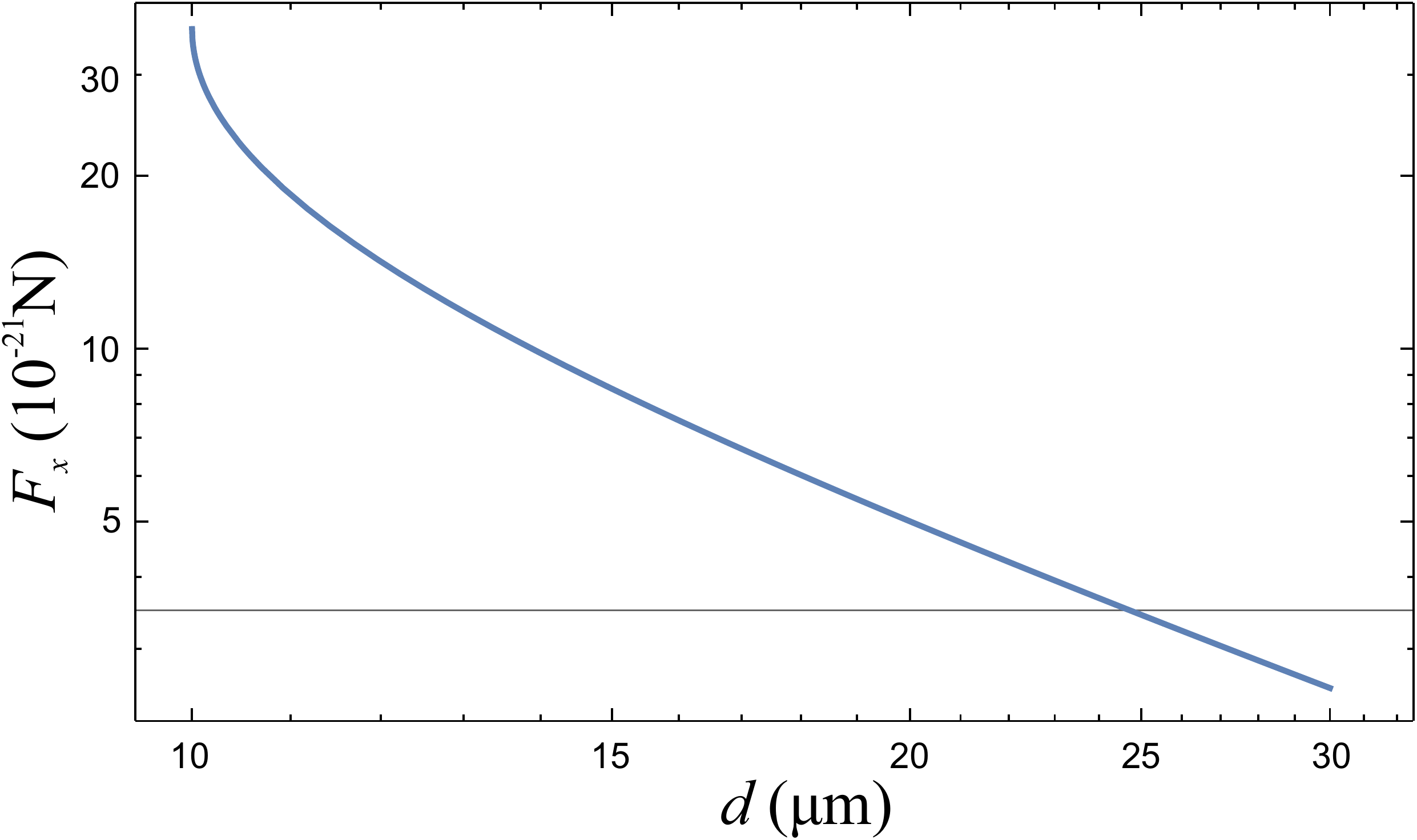}
\caption{ Force acting on the magnet due to magnetic field fluctuations as a function of the distance to the plane $d$ evaluated at $\omega/(2\pi)\approx 25$Hz. The force sensitivity threshold $\sqrt{S_{F_x}}\approx 3.5\times 10^{-21}$N/$\sqrt{\rm Hz}$ is indicated with a gray line. }
\label{fig.Cas}
\end{center}
\end{figure}

\textbf{\textit{Magnetic fields}}. Gradients of external magnetic fields result in forces acting on the magnet. Considering the magnet as a point particle, with a magnetic moment that points to the $y$-direction, these forces read $F_i=\mu\, \partial_i B_y$. The PSD of the position signal resulting from these forces is $S_{r_i}^{\rm B}=|\chi_{ii}\, \mu|^2S_{\partial_iB_y}$ and the PSD of the corresponding force is thus
\begin{equation}
S_{F_i}^{\rm B}=|\mu|^2S_{\partial_iB_y}.
\end{equation}
When the source of magnetic field is near to the magnet, the point-particle approximation may not be valid. In this case the total force acting on the magnet can be calculated as an integral over its surface $S$
\begin{equation}
\mathbf{F}=\int_S \mathbf{K}_{\rm M}(\mathbf{r}')\times\mathbf{B}(\mathbf{r}') {\rm d}V',\label{eq.integ}
\end{equation}
where $\mathbf{K}_{\rm M}\equiv \nabla \times \mathbf{M}$ is the magnetization sheet current density. For the case of magnetic fields arisen from fluctuating currents in a neutral surface (magnetic Casimir forces) \cite{henkel1999}, these magnetic field fluctuations are 
\begin{equation}
B^{\rm M}_i (d,\omega)\approx\sqrt{\frac{\mathcal{C}(\omega)}{d}}
\end{equation}
where $d$ is the distance to the surface, $i=\{x,y,z\}$, and $\mathcal{C}(\omega)=\mu_0^2 \omega^2 \hbar \varepsilon_0 {\rm Im}[\varepsilon(\omega)]/(16 \pi)$, where $\epsilon(\w)$ is the dielectric constant in the spectral representation of the magnetic source, and $\epsilon_0$ the vacuum permittivity. Consider a surface parallel to the plane ZY at $x=d$. The magnet will experience a force in $x$-direction given by the surface integral of Eq.~\eqref{eq.integ}, namely
\begin{equation}
F_x(d,\omega)=R^2 \frac{\mu}{V}\int_0^{2\pi} {\rm d}\phi\int_0^{\pi} {\rm d}\theta  B_y^{\rm M} \sin^2 \theta \cos \phi, 
\end{equation}
with $B_y^{\rm M}=\sqrt{\mathcal{C}(\omega)/(d-R \sin\theta \cos\phi)}$. The force is evaluated using the low-frequency limit for the dielectric function, ${\rm Im}[\varepsilon(\omega)]=1/(\varepsilon_0 \omega \rho)$, being $\rho$ the electric resistance of the surface~\cite{henkel1999}. For a magnet of $R=10\mu$m and a surface made of silver with $\rho=1.6\times10^{-8}\Omega$m the force as a function of the distance $d$ is shown in Fig.~\ref{fig.Cas} [at a frequency $\omega/(2\pi)\approx 25$Hz]. Considering that the force sensitivity for this size of magnet at this same frequency is $\sqrt{S_{F_x}}\approx 3.5\times 10^{-21}$N/$\sqrt{\rm Hz}$, forces fall within the detectability threshold for distances up to $d\approx25\mu$m, corresponding to separations of around $15\mu$m from the magnet. 

\section{Readout and feedback cooling}

For the readout system we consider 4 identical adjacent square loops in a XZ plane. We label them through the position of their centers $\mathbf{r}^c$: 
(1) $z^c_{1}>z_0$ and $x^c_{1}>0$, 
(2) $z^c_{2}<z_0$ and $x^c_{2}>0$,
(3) $z^c_{3}>z_0$ and $x^c_{3}<0$, 
(4) $z^c_{4}<z_0$ and $x^c_{4}<0$. 
Taking into account Eq.~\eqref{eq.fl} we can now write an equation for each loop
\begin{align}
\begin{split}
\Delta \Phi_1(\mathbf{r})&=\Phi_0  (+\eta_x x-\eta_y y+\eta_z z),\\
\Delta \Phi_2(\mathbf{r})&=\Phi_0  (+\eta_x x-\eta_y y-\eta_z z),\\
\Delta \Phi_3(\mathbf{r})&=\Phi_0  (-\eta_x x-\eta_y y+\eta_z z),\\
\Delta \Phi_4(\mathbf{r})&=\Phi_0  (-\eta_x x-\eta_y y-\eta_z z),
\end{split}
\end{align}
where $\Delta \Phi(\mathbf{r})=\Phi(\mathbf{r}_0+\mathbf{r})-\Phi(\mathbf{r}_0)$ corresponds to the variation of magnetic flux measured by the $i$th loop. Notice that the absolute value of the coupling factors is the same for all the loops due to their symmetric arrangement. Only their signs change. The position of the magnet can be then determined by solving this system of equations. Also notice that the signal of the four loops is added up to determine the position of the magnet. For this reason, the coupling factors provided in the next section already contain the contribution of the four loops.  

\begin{figure}[t]
\begin{center}
\includegraphics[width=0.50\textwidth]{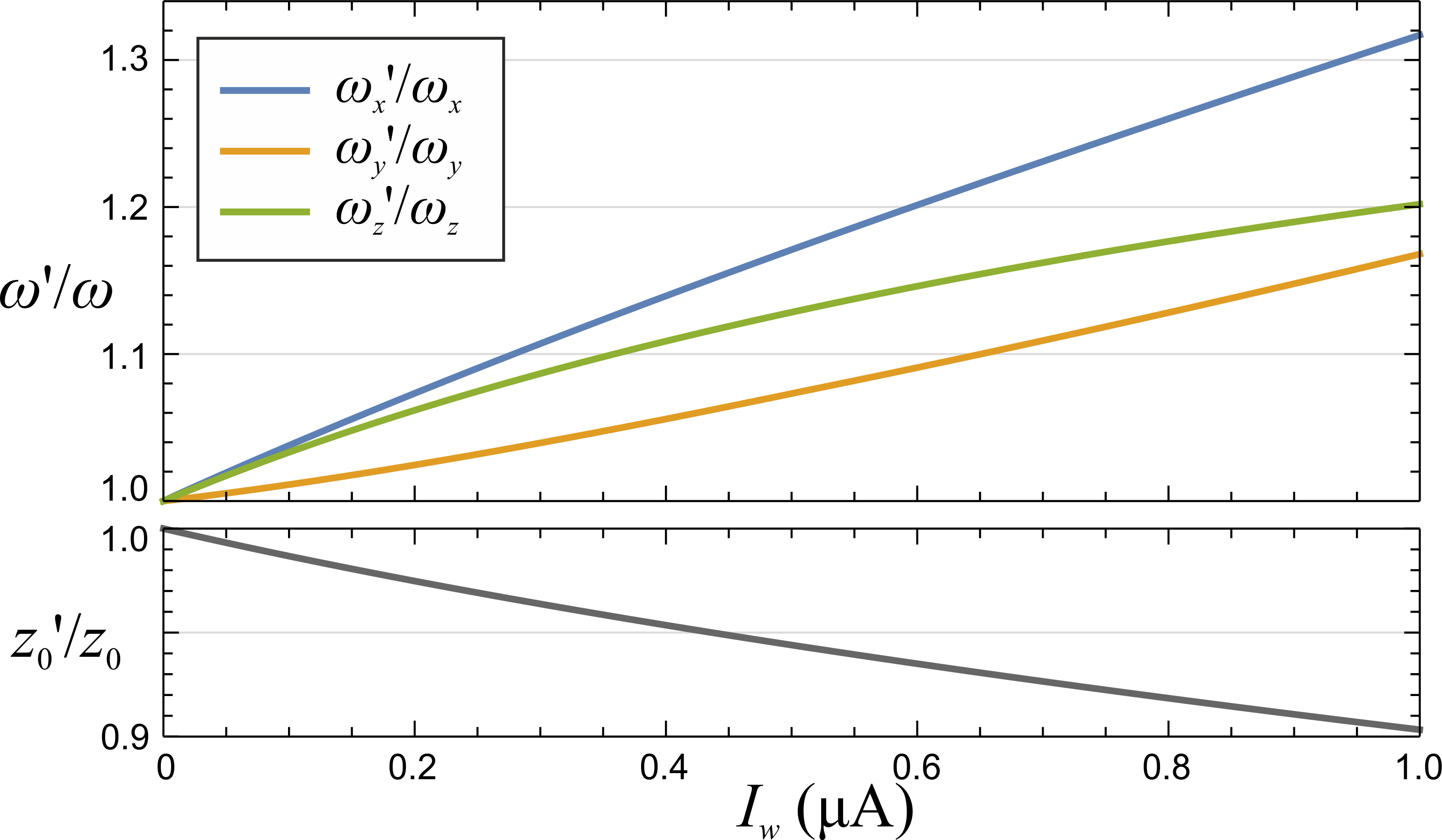}
\caption{Change in the trapping frequencies (upper half) and trapping position of the magnet (lower half) as a function of the intensity in the wire.}
\label{eq.fe}
\end{center}
\end{figure}

We now show how the trapping position of the magnet and the trapping frequencies can be modified by feeding current to a wire parallel to the $x$-axis at a given $z_{\rm w}$. We consider a particular example with $R=0.1\mu$m, for which the magnet is trapped at $z_0\approx 5\mu$m. The wire is set at $z_{\rm w}=8\mu$m and a given intensity $I_{\rm w}$ circulates in the direction defined by $-\mathbf{e}_x$. As shown in Fig.~\ref{eq.fe}, both the trapping position and the frequencies are modified by changing the intensity in the wire. This modulation could be used to perform parametric feedback cooling of the levitated magnet. A thorough analysis on how to perform it in an optimal way such that the added noise does not compromise the overall sensitivity will be addressed elsewhere.

\section{Study case}

\begin{figure}[b]
\begin{center}
\includegraphics[width=0.50\textwidth]{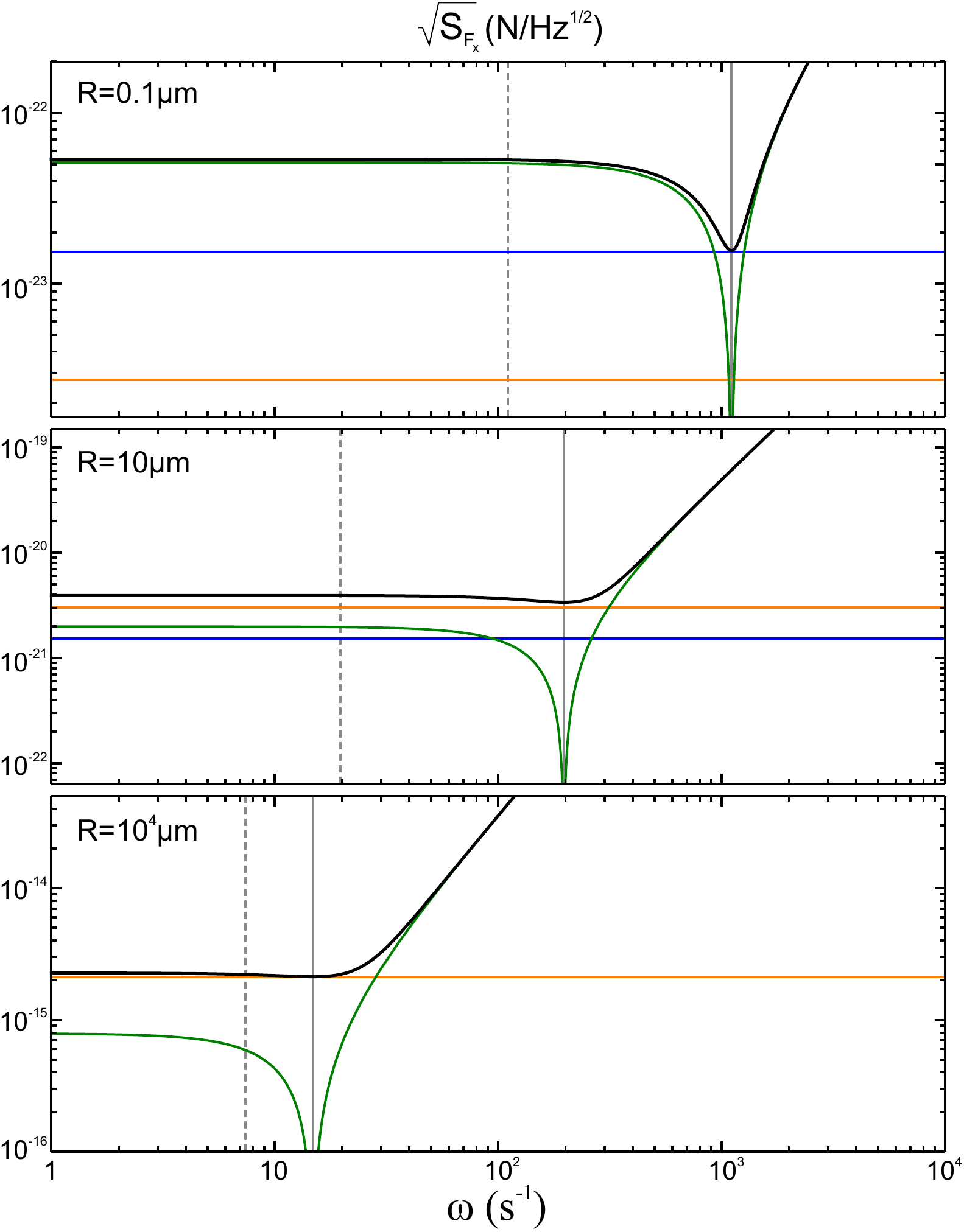}
\caption{Plots of the noises ($x$-component of the force) for different radii of the magnet as a function of the frequency. From top to bottom, $R=0.1$, $10$ and $10^4\mu$m. The color legend is the same as in Fig.~2 of the main text; $\sqrt{S_{F_x}^{\rm g}}$ in blue, $\sqrt{S_{F_x}^{\rm h}}$ in orange, $\sqrt{S_{F_x}^{\rm S}}$ in green, and total $\sqrt{S^{\star}_{F_x}}$ in black. Vertical solid gray lines indicate the corresponding resonance frequencies ($\omega_x$) and dashed lines are the frequencies where sensitivities have been evaluated ($0.1\omega_x$ for the first two cases and $0.5\omega_x$ for the latter).}
\label{plots.fre}
\end{center}
\end{figure}

Noise results presented in Fig.~2 of the main text have been calculated assuming the material parameters of Nd$_2$Fe$_{14}$B \cite{coey}. We used $\mu=\rho_{\mu} V$ and $M=\rho_{M} V$, with $V$ being the volume of the magnet and $\rho_{\mu}=1.07\times 10^6$A/m and $\rho_{M}=7300$Kg/m$^3$. We also considered a magnetic susceptibility $\chi_{\rm m}=0.05$ and an electrical conductivity $\sigma=6.67\times10^5$A/(V$\cdot$m). For the environment, we considered a pressure of $P=10^{-10}$mbar of a gas with molar mass of $28.97$u at a temperature of $T=1$K. For the SC sheet we assume it to be made of Niobium with a critical temperature $T_c=9.26$\,K. Below the first critical field $H_{c1}$, Nb behaves as a superconductor in the Meissner state provided it is cooled in zero-field. The normalized trapping position of the magnet is always set to $z_0/a=1.8$ (as in Fig.~1 of the main text). For the readout system of SQUIDS, we adapted the distance to the magnet and their size as a function of the radius of the magnet. For simplicity, we considered four identical adjacent square loops of side length $s$. They were placed on the same plane, parallel to the plane XZ at a distance $d_h$ and with centers at positions $(\pm s/2,-d_h,z_0\pm s/2)$. The side length of the loops was chosen such that the three coupling factors have similar values $\eta_x\approx \eta_y \approx \eta_z$. Table \ref{table} summarizes the parameters. 
\begin{table}[h!]
\begin{tabular}{c |c| c| c}
R ($\mu$m)              & $d_h$ ($\mu$m) & $s$ ($\mu$m) & $\eta$ (m$^{-1}$)  \\
\hline
0.1          & 1 & 0.85 & $7.3\times10^5$  \\
1            & 4 & 3.5 & $4.6\times10^7$  \\
10           & 20 & 17 & $1.8\times10^9$  \\
$10^{2}$     & 110 & 95 & $6.0\times10^{10}$  \\
$10^{3}$     & 1100 & 950 & $6.0\times10^{11}$  \\
$10^{4}$     & 11000 & 9500 & $6.0\times10^{12}$  \\
\hline
\end{tabular}
\caption{Summary of the parameters for the readout system.}
\label{table}
\end{table}
Finally, for a radius of the magnet between $0.1$ and $100\mu$m we evaluated the noises at a frequency of $\omega= 0.1 \omega_i$, being $\omega_i$ the corresponding resonance frequency. For radii bigger than $100\mu$m, noises were evaluated at $\omega= 0.5 \omega_i$. In \figref{plots.fre} force noises ($x$-component) are plotted for three different radii of the magnet as a function of the frequency. Vertical dashed lines indicate the frequency at which noises have been evaluated to make Fig.~2 of the main text.

\end{document}